\documentclass[journal=jacsat,manuscript=article]{achemso}
\PassOptionsToPackage{sort&compress,numbers,super}{natbib}

\usepackage{graphicx}
\usepackage{amsmath}
\usepackage{amssymb}
\usepackage{textcomp}
\usepackage{siunitx}
\usepackage{booktabs}
\usepackage{float}
\usepackage{mhchem}
\usepackage{hyperref}
\usepackage{hyperref}
\usepackage[capitalise,noabbrev]{cleveref}

\usepackage[T1]{fontenc}
\usepackage[utf8]{inputenc}
\usepackage{lmodern}
\usepackage{textcomp}

\title{Vacancy-Induced Quantum Properties in 2D Silicon Carbide: Atomistic insights from semi-local and hybrid DFT calculations}

\author{Abhirup Patra}
\affiliation{Department of Chemical and Biomedical Engineering, University of Delaware}
\alsoaffiliation{Delaware Energy Institute, University of Delaware, 221 Academy St, Newark, DE 19711, USA}
\email{patraa@udel.edu}

\begin{document}
	
	\begin{abstract}
		Two-dimensional (2D) materials have emerged as promising platforms for quantum technologies and optoelectronics, with defects playing a crucial role in their properties. We present a comprehensive density functional theory study of silicon and carbon vacancies in monolayer silicon carbide (1L-SiC), a wide-bandgap 2D semiconductor with potential for room temperature quantum applications. Using PBE, SCAN, r$^2$SCAN, and HSE06 functionals, we reveal distinct characteristics between Si and C vacancies. Formation energies and charge transition levels show strong functional dependence, with HSE06 consistently predicting higher values and deeper transition levels compared to PBE calculations. Electronic structure analysis demonstrates contrasting behavior: silicon vacancies create highly localized states with strong spin polarization, while carbon vacancies produce more dispersed states with weaker magnetic properties. Vacancy migration studies reveal significantly lower barriers for silicon vacancies compared to carbon vacancies, indicating higher mobility for Si vacancies at moderate temperatures. Optical properties, calculated using PBE-DFPT, show distinct charge-state dependent absorption in the far-infrared region, with positively charged states of both vacancy types demonstrating the strongest response. The complementary characteristics of Si and C vacancies-localized versus dispersed states, different magnetic properties, and distinct optical responses-suggest possibilities for defect engineering in quantum and optoelectronic applications. Our results highlight the critical importance of advanced functionals in accurately describing defect properties and provide a comprehensive framework for understanding vacancy behavior in 2D materials.
	\end{abstract}
	
	\section{Introduction}
	
	\textcolor{blue}{The field of two-dimensional (2D) semiconductors has witnessed remarkable progress in recent years, driven by the discovery of novel materials and their intriguing properties \cite{Falk2015,Nagy2019,Wolfowicz2021}. Following the initial isolation of graphene, other 2D materials, including transition metal dichalcogenides (TMDs), hexagonal boron nitride (h-BN), and phosphorene, have been extensively investigated. These atomically thin materials, characterized by their unique electronic, optical, and mechanical properties, have opened exciting avenues for applications, particularly in quantum information science and optoelectronics \cite{Baranov2011,Riedel2012,Szasz2013,Castelletto2014,Christle2015,Radulaski2017,Davidsson2018}. Among these emerging 2D semiconductors, silicon carbide (SiC) has attracted significant attention due to its exceptional quantum properties that persist even at room temperature \cite{Castelletto2014,Falk2015,Nagy2019}. Defects within SiC, such as silicon vacancies (V$_\text{Si}$) and carbon antisites (Si$_\text{C}$), play a pivotal role in enabling single-photon emission, a critical feature for quantum communications \cite{Falk2015,Nagy2019,Wolfowicz2021}. Through advanced spectroscopic techniques like electron spin resonance (ESR) and photoluminescence (PL), researchers have gained valuable insights into the nature of these defect states and their impact on the material's electronic and optical properties \cite{Baranov2011,Riedel2012,Szasz2013,Castelletto2014,Christle2015,Radulaski2017,Davidsson2018}. Theoretical predictions suggest that local defects in 2D SiC could offer superior photon emission efficiency due to reduced dielectric screening and enhanced quantum confinement effects \cite{Gali2019,Udvarhelyi2020}.}
	The single-atomic-layer thickness of 2D SiC minimizes the path length for emitted photons, potentially increasing extraction efficiency compared to bulk materials where total internal reflection can trap emitted photons. Additionally, the reduced dimensionality enhances the oscillator strength of optical transitions, potentially leading to brighter single-photon sources. \textcolor{blue}{Furthermore, the existence of Stone-Wales defects \cite{stone1986theoretical,Pizzochero2019}, created by rotating a carbon-carbon bond in the SiC lattice, has been proposed as offering superior radiative transitions compared to isolated point defects due to their unique electronic structure.} Despite these promising theoretical predictions, realizing the potential of 2D SiC in practice remains a significant challenge due to difficulties in synthesizing high-quality 2D SiC and controlling defects at the atomic level.\cite{islam2020vacancy,fan2006low,Mishra2016,Volodin2018,chabi2021creation} Current synthesis approaches include both top-down methods such as selective etching and mechanical exfoliation, and bottom-up techniques like chemical vapor deposition and molecular beam epitaxy.\cite{yamasue2015interfacial,Nilius2002} However, these methods often result in samples with high defect concentrations or limited size, hindering their practical applications. Controlling defect type, concentration, and spatial distribution remains a critical challenge that must be addressed to harness the quantum properties of 2D SiC effectively. To characterize and study the defects in 2D SiC, advanced techniques with atomic-scale resolution are essential. Scanning tunneling microscopy (STM),\cite{Nilius2002} high-resolution transmission electron microscopy (HRTEM), and specialized optical characterization methods like photoluminescence (PL) spectroscopy have been instrumental in probing the atomic structure and properties of defects.\cite{Castelletto2014,fuchs2015engineering,Manson2016,Bockstedte2018,Koehl2018,Wang2020} Substantial progress has been achieved in comprehending the dynamics of defects in 2D SiC via advanced theoretical research, particularly with respect to thermodynamic stability.\cite{Pizzochero2019,Udvarhelyi2020,Abellan2021} \textcolor{blue}{First-principles calculations based on density functional theory (DFT) have provided valuable insights into the formation energies, electronic structure, and optical properties of various defects in 2D SiC \cite{Udvarhelyi2020,Pizzochero2019,Abellan2021,ha2023novel}.} By comparing the calculated defect formation energies and charge transition levels with experimental data, researchers can identify the most stable defect configurations and predict their behavior under different environmental conditions. Pizzochero and Berthod\cite{Pizzochero2019} conducted a comprehensive DFT study examining the energetic stability of various defects in monolayer SiC, finding that the divacancy complex (V$_\text{Si}$-V$_\text{C}$) exhibits remarkably low formation energy and is thermodynamically favorable compared to isolated vacancies. This finding has important implications for experimentalists aiming to engineer and control defects in freestanding 2D SiC, as the divacancy-complex can serve as a stable and optically active center for quantum applications. Moreover, DFT calculations have provided valuable insights into the spin-strain coupling properties of defects in 2D SiC.\cite{Udvarhelyi2020} By analyzing the electronic structure and spin density distribution of defects under different strain conditions, researchers can predict the sensitivity of defect-related optical transitions to mechanical deformation. This knowledge is crucial for developing strain-tunable quantum emitters and sensors based on 2D SiC. Udvarhelyi et al.\cite{Udvarhelyi2020} demonstrated through DFT calculations that certain defect complexes in 2D SiC exhibit strong spin-strain coupling, potentially enabling mechanical control of their quantum states. Recent theoretical studies have also explored the potential of defect engineering in 2D SiC through the incorporation of foreign atoms or the creation of specific defect complexes.\cite{Abellan2021} DFT calculations have predicted that doping 2D SiC with elements such as nitrogen or boron can introduce new energy levels and optical transitions, enabling the tuning of the material's electronic and optical properties. Recent experimental breakthroughs in synthesizing 2D SiC have validated many theoretical predictions about its properties and stability.\cite{chabi2021creation,polley2023bottom} Both top-down chemical exfoliation\cite{chabi2021creation} and bottom-up epitaxial growth on transition metal carbides\cite{polley2023bottom} have demonstrated the feasibility of producing monolayer SiC. These advances provide crucial experimental validation of the structural and electronic properties predicted by theory, while also revealing new phenomena such as strong spin-orbit coupling effects in substrate-supported 2D SiC.\cite{polley2023bottom} However, the accuracy of DFT calculations hinges critically on the chosen exchange-correlation functional. The Perdew-Burke-Ernzerhof (PBE) functional offers an efficient approximation but often falls short in capturing the correct electronic structure and defect formation energies.\cite{Perdew1996,Zhang1998,Adamo1999} PBE tends to underestimate the band gap of semiconductors and overdelocalize defect states, leading to an underestimation of defect formation energies and charge transition levels. This limitation can be particularly problematic for defects in wide-bandgap materials like SiC, where accurate description of localized defect states is crucial. The self-interaction error inherent in semi-local functionals like PBE arises from the incomplete cancellation of the electron self-interaction in the exchange-correlation energy. This error is particularly pronounced for localized states, such as those introduced by defects in semiconductors, leading to an artificial delocalization of these states and consequently incorrect predictions of their energetics and electronic structure. The underestimation of bandgaps in semiconductors by PBE is a direct consequence of this self-interaction error, which affects both the valence and conduction band edges but typically lowers the conduction band more significantly, resulting in reduced bandgaps compared to experimental values. Hybrid functionals like HSE06 have emerged as a more reliable choice for defect calculations in semiconductors.\cite{heyd2003hybrid,krukau2006influence,paier2007does} By incorporating a fraction of exact exchange from Hartree-Fock theory \cite{fock1930naherungsmethode,lykos1963discussion}, HSE06 can partially correct the self-interaction error and improve the description of localized defect states. HSE06 has been shown to provide accurate band gaps and defect formation energies for a wide range of semiconductor materials, including 2D SiC. The mixing of exact exchange with semi-local exchange in HSE06 is particularly effective at correcting the self-interaction error for localized states, resulting in more accurate predictions of defect formation energies, charge transition levels, and magnetic properties. \textcolor{blue}{However, the computational cost of HSE06 calculations, scaling as $O(N^4)$ due to the inclusion of exact exchange, is significantly higher than that of PBE, which scales as $O(N^3)$, limiting HSE06’s applicability to large-scale simulations.} To bridge the gap between accuracy and computational efficiency, meta-GGA functionals like SCAN and r$^2$SCAN have been developed.\cite{Sun2015,Sun2016,Furness2020,Zhang2021} These functionals go beyond the local density approximation (LDA) and GGA by incorporating higher-order derivatives of the electron density, such as the kinetic energy density. SCAN and r$^2$SCAN have shown promising performance in describing the electronic structure and thermodynamic properties of a wide range of materials, including semiconductors and 2D systems. Recent studies have demonstrated that SCAN and r$^2$SCAN can provide accurate band gaps and defect formation energies for 2D SiC, comparable to those obtained with HSE06.\cite{Furness2020,Zhang2021} The strongly constrained and appropriately normed (SCAN) meta-GGA functional was designed to satisfy all known exact constraints for exchange-correlation functionals and is appropriately normed to accurate reference data for systems where the exact functional should recover the exact energy.\cite{Sun2015} By incorporating the kinetic energy density, SCAN can distinguish between different chemical bonding environments, leading to improved descriptions of both covalent bonds and weak interactions. This is particularly beneficial for materials like SiC, which feature a mix of covalent and ionic bonding characters. However, SCAN has been found to be numerically challenging to converge in some cases, leading to the development of the regularized SCAN (r$^2$SCAN) variant, which maintains most of SCAN's accuracy while offering improved numerical stability.\cite{Furness2020} The computational cost of SCAN and r$^2$SCAN calculations is significantly lower than that of HSE06, making them attractive options for large-scale simulations of defects in 2D materials. While these functionals offer improved accuracy compared to PBE, they may still not fully capture the exact exchange effects necessary for highly localized defect states, particularly those with strong magnetic character. Therefore, a careful comparison of different functionals is essential for establishing reliable computational protocols for studying defects in 2D SiC. The objective of this research is to systematically examine the formation energies, electronic properties, migration barriers, and optical responses of Si and C vacancies in monolayer SiC using multiple DFT functionals (PBE, SCAN, r$^2$SCAN, and HSE06). By comparing these functionals, we aim to evaluate their effectiveness in representing localized defect states and to comprehend the impact of these states on the material's quantum characteristics. Our comprehensive approach includes accurate charge correction schemes for charged defects and detailed analysis of spin polarization effects. The study aims to provide valuable insights into the theoretical basis of SiC's defects and aid in selecting appropriate computational strategies in the rapidly expanding field of SiC-based technologies.
	
	\section{Computational Methods}
	We conducted spin-polarized density functional theory (DFT) calculations using the Vienna Ab initio Simulation Package (VASP)\cite{kresse1996efficient} with plane-wave basis sets, PAW-PBE pseudopotentials,\cite{blochl1994projector} and an energy cutoff of 450 eV. \textcolor{blue}{This energy cutoff was chosen after convergence testing, which showed that increasing the cutoff beyond 450 eV changed total energies by less than 1 meV per atom. (see Supplementary Figure S1 for convergence graphs) We employed four different exchange-correlation functionals to systematically evaluate their performance in describing defect properties: (1) the Generalized Gradient Approximation (GGA) with Perdew-Burke-Ernzerhof (PBE),\cite{Perdew1996} (2) the strongly constrained and appropriately normed (SCAN) meta-GGA functional,\cite{Sun2015} (3) its modified version r$^2$SCAN,\cite{Furness2020} and (4) the Heyd-Scuseria-Ernzerhof (HSE06) hybrid exchange-correlation functional,\cite{heyd2003hybrid} all including DFT-D3 van der Waals corrections.\cite{caldeweyher2017extension} Complete relaxation of atomic positions and supercell dimensions was allowed using conjugate gradient optimization techniques without symmetry constraints. Force convergence criteria of 0.02 eV/$\AA$ and total energy precision within 0.01 meV were achieved with Brillouin zone sampling on a 2×2×1 Monkhorst-Pack k-point grid for structural optimizations and a denser 4×4×1 grid for band structure calculations.\cite{monkhorst1976special} This k-point sampling was verified to be sufficient through convergence tests that showed energy differences less than 5 meV when compared to denser grids (Supplementary Figure S2). We employed a 3×3 supercell (containing 18 atoms for pristine SiC) for simulating the defective systems. This supercell size was chosen as a balance between computational cost and accuracy, and we performed convergence tests with larger 4×4 and 5×5 supercells for selected configurations to verify that our 3×3 supercell results are well-converged. Energy differences between the 3×3 and 5×5 supercells were found to be less than 0.05 eV for formation energies (Supplementary Figure S3) and less than 0.1 eV for migration barriers, confirming the adequacy of our chosen supercell size. The initial crystal structure of monolayer SiC was obtained from theoretical models of the hexagonal 2D SiC lattice, as reported in previous studies \cite{ha2023novel,lu2012tuning}. A 3×3 supercell containing 18 atoms (9 Si and 9 C) was constructed from this pristine structure. Vacancies were introduced by removing a single Si or C atom from the supercell, followed by full structural relaxation to account for local distortions around the defect site.} A vacuum length of 20 $\AA$ in the z-direction was used to prevent spurious interactions between periodic images. This vacuum size was determined through convergence testing, which showed negligible energy changes ($<$ 1 meV) when increasing the vacuum beyond 20 $\AA$. For the electronic structure calculations, we employed the tetrahedron method with Blöchl corrections for Brillouin zone integration to obtain accurate density of states (DOS) and band structures. The DOS calculations utilized a denser 6×6×1 k-point grid to ensure smooth spectra, while band structures were calculated along the high-symmetry path $\Gamma-M-K-\Gamma$ in the Brillouin zone. For charged defect calculations, we included a compensating homogeneous background charge to maintain overall charge neutrality in the periodic system, as is standard practice in plane-wave DFT calculations.
	
	We determined vacancy defect formation energies in monolayer SiC by applying thermodynamic principles and considering carbon-rich conditions. The formation energies of defects in charge-state $q$ were calculated using the equation:
	
	\begin{equation}
		\Delta E_f(q) = E_\text{tot}(q) - E_\text{bulk} - \sum_A n_A\mu_A + qE_\text{Fermi} + E_\text{corr}
	\end{equation}
	
	Here, $E_\text{tot}(q)$ is the total energy of the supercell containing one defect of charge state $q$, $E_\text{bulk}$ is the total energy of the defect-free supercell, $\mu_A$ is the chemical potential of the removed atom type, $n_A$ is the number of missing atoms, $E_\text{Fermi}$ is the Fermi energy referenced to the theoretical valence band maximum (VBM) for a given functional, and $E_\text{corr}$ is the correction term for electrostatic interactions between periodically repeated defects. The chemical potentials of the constituent elements are subject to thermodynamic constraints. In this work, $\mu_\text{Si}$ is derived from diamond silicon, while $\mu_\text{C}$ is calculated from bulk graphite. This approach corresponds to carbon-rich conditions, which are likely to be encountered in experimental synthesis of 2D SiC. Specifically, we calculated the bulk energies of diamond silicon and graphite, and used these values as the respective chemical potentials, ensuring that $\mu_\text{Si} + \mu_\text{C} = \mu_\text{SiC}$ (bulk) to maintain thermodynamic equilibrium with bulk SiC. These chemical potential references were chosen to represent realistic experimental conditions, though the absolute formation energies would differ under silicon-rich conditions. For charged defect calculations, we employed two different charge correction schemes to account for the spurious electrostatic interactions that arise in periodic supercell approaches: the Freysoldt, Neugebauer, and Van de Walle (FNV) method\cite{freysoldt2009fully} and the Kumagai \& Oba (KO) method.\cite{kumagai2014electrostatics} These corrections are essential for obtaining accurate formation energies of charged defects, as the periodic boundary conditions used in plane-wave DFT calculations can lead to significant errors due to the long-range nature of Coulomb interactions.
	
	The FNV method addresses the spurious electrostatic interaction between periodic images of charged defects in a supercell using an isotropic dielectric constant:
	
	\begin{equation}
		E_\text{corr}^\text{Freysoldt} = E_f + E_\text{corr}(q, \varepsilon, L) = E_\text{image} + V_\text{align}
	\end{equation}
	
	where $q$ is the charge of the defect, $\varepsilon$ is the dielectric constant, $L$ is the supercell dimension, $E_\text{image}$ is the image charge correction term, and $V_\text{align}$ is the potential alignment term. The KO method extends the Freysoldt approach by including the anisotropic nature of the dielectric response, which is particularly important for 2D materials:
	
	\begin{equation}
		E_\text{corr}^\text{Kumagai} = E_f + E_\text{corr}(q, \varepsilon_\text{ani}, L) = E_\text{image,ani} + V_\text{align,ani}
	\end{equation}
	
	where $\varepsilon_\text{ani}$ is the anisotropic dielectric tensor. This approach accounts for the different screening responses in the in-plane and out-of-plane directions, providing more accurate corrections for charged defects in 2D systems. To obtain the dielectric properties needed for these corrections, we calculated the static dielectric constant $\varepsilon$ and Born effective charges using density functional perturbation theory (DFPT) as implemented in VASP. For these calculations, we used an 8×8×1 k-point mesh on a 3×3 supercell with the PBE-GGA functional. The computed dielectric constants of pristine 1L-SiC were $\varepsilon_{||} = 4.2$ for the in-plane component and $\varepsilon_{\perp} = 1.8$ for the out-of-plane component, yielding an isotropic average of $\varepsilon = 9.3$, which was used in the FNV correction scheme. For the KO correction method, we employed the full dielectric tensor to account for the anisotropic screening effects in the 2D material.
	
	Migration pathways and diffusion barriers ($E_b$) of defect sites were computed using the climbing-image nudged elastic band (CI-NEB) method,\cite{henkelman2000improved} as presented in Figure 5. This method provides an efficient way to find minimum energy paths between known initial and final states, and accurately determines the transition state energy and configuration. We created seven intermediate images to sample the minimum energy path between the initial and final configurations, which was sufficient to obtain smooth energy profiles and accurately locate transition states. The initial and final states for each migration pathway were fully relaxed before constructing the NEB path. For each vacancy type (Si and C), we considered three distinct migration mechanisms, labeled as Paths A, B, and C, representing different possible routes for vacancy diffusion in the 2D SiC lattice. Path A corresponds to migration to the nearest equivalent site within a single hexagon, Path B involves migration across two hexagons, and Path C represents migration to the nearest neighbor of the same atom type. The spring constant between images was set to 5.0 eV/$\AA$², which provided adequate coupling between images while allowing sufficient flexibility for the path to converge to the minimum energy route. The climbing image algorithm was employed to precisely locate the transition state by maximizing its energy along the reaction coordinate while minimizing it in all other directions. Force convergence for the NEB calculations was set to 0.05 eV/$\AA$, which is slightly less stringent than for structural relaxations but sufficient for accurate barrier determination.
	
	For the optical properties, we calculated the frequency-dependent dielectric function $\varepsilon(\omega)$ within the independent particle approximation using DFPT. This approach, while neglecting excitonic effects, has been shown to provide reliable trends in optical responses for defect systems.\cite{Gali2019} The imaginary part $\varepsilon_2(\omega)$ was obtained by summing over empty states according to:
	
	\begin{equation}
		\varepsilon_2(\omega) = \frac{4\pi^2e^2}{\Omega}\lim_{q\to 0}\frac{1}{q^2}\sum_{c,v,k}2w_k\delta(\varepsilon_{ck}-\varepsilon_{vk}-\hbar\omega)|\langle u_{ck+e_\alpha q}|u_{vk}\rangle|^2
	\end{equation}
	
	where $c$ and $v$ refer to conduction and valence band states respectively, $u_{ck}$ are the cell periodic parts of the wavefunctions at point $k$, $w_k$ are the k-point weights, and $\Omega$ is the cell volume. The real part $\varepsilon_1(\omega)$ was obtained using the Kramers-Kronig relations:
	
	\begin{equation}
		\varepsilon_1(\omega) = 1 + \frac{2}{\pi}P\int_0^\infty \frac{\omega'\varepsilon_2(\omega')}{\omega'^2-\omega^2}d\omega'
	\end{equation}
	
	where $P$ denotes the principal value of the integral. From $\varepsilon_1(\omega)$ and $\varepsilon_2(\omega)$, we calculated the frequency-dependent absorption coefficient:
	
	\begin{equation}
		\alpha(\omega) = \frac{2\omega}{c}\sqrt{\frac{\sqrt{\varepsilon_1^2(\omega) + \varepsilon_2^2(\omega)} - \varepsilon_1(\omega)}{2}}
	\end{equation}
	
	where $c$ is the speed of light. The optical properties were calculated using DFPT within the PBE framework, which provides a computationally efficient approach for excited state properties while maintaining good accuracy for relative changes in optical response. To ensure accuracy of our optical calculations, we employed a dense 12×12×1 k-point mesh, significantly finer than that used for ground state properties. The calculations included 100 empty bands, chosen after convergence testing showed variations below 0.1\% in the computed dielectric function when increasing to 120 bands. We applied a Gaussian broadening of 0.1 eV to the optical spectra, balancing spectral resolution with numerical stability. Local field effects were fully incorporated in our calculations through the inclusion of off-diagonal elements of the dielectric matrix up to 100 eV in the plane-wave basis. The frequency-dependent dielectric response was computed up to 5 eV using a dense frequency grid of 0.02 eV to accurately capture fine spectral features. The absorption coefficients ($\alpha$) were computed from the complex dielectric function and are presented as percentages relative to the pristine case to facilitate direct comparison of vacancy effects. The energy range was chosen to focus on the far-infrared region (0-0.5 eV) where vacancy-induced features are most prominent.
	
	\section{Results and Discussion}
	
	\subsection{Defect Formation Energies}
	
\begin{figure}[H]
		\centering
		\includegraphics[width=1.2\textwidth]{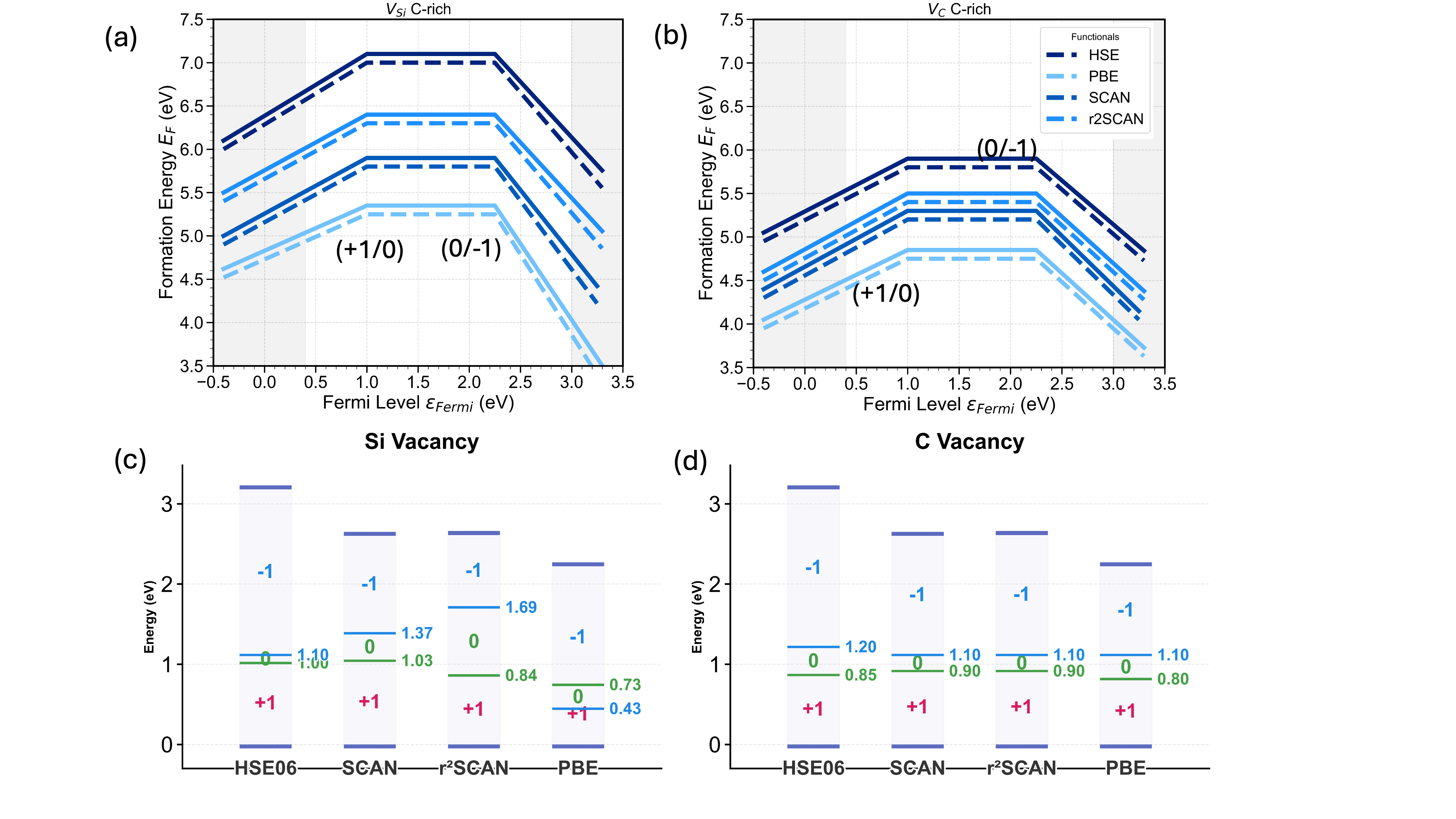}
		\caption{Comparison of defect formation energies for (a) Si vacancy and (b) C vacancy in C-rich environments. The FNV charge correction scheme results are displayed by the dashed line, while the KO charge correction scheme results are shown by the solid line. The range of Fermi levels is calculated based on the thermodynamic stability region in C-rich conditions. The slopes of +1 and -1 for the +1 and -1 charge states respectively reflect the physical charge transfer process.}
		\label{fig:formation}
	\end{figure}

	Figure \ref{fig:formation} presents the analysis of vacancy formation energies in monolayer SiC under carbon-rich conditions, calculated using a $3\times3\times1$ supercell with convergence testing performed up to $5\times5\times1$ cells. The formation energies are plotted as a function of the Fermi level ($E_{Fermi}$) for both silicon vacancies (Fig. \ref{fig:formation}(a)) and carbon vacancies (Fig. \ref{fig:formation}(b)). These energies were calculated using four distinct exchange-correlation functionals: PBE, SCAN, r$^2$SCAN, and HSE06. The figure displays results obtained using two charge correction schemes: the Freysoldt-Neugebauer-Van de Walle (FNV) method\cite{freysoldt2009fully}, shown with dashed lines, and the Kumagai-Oba (KO) method\cite{kumagai2014electrostatics}, represented by solid lines. The range of Fermi levels shown corresponds to the thermodynamic stability region under C-rich conditions.
	
	For silicon vacancies (Fig. \ref{fig:formation}(a)), the formation energies calculated at the valence band maximum (VBM, $E_{Fermi} = 0$) show a pronounced dependence on the chosen functional. Using the KO correction scheme (solid lines), the energies range from approximately 4.3 eV for PBE (lightest blue) to about 6.0 eV for HSE06 (darkest blue). This significant variation of approximately 2.8 eV underscores the influence of the exchange-correlation functional on the calculated defect energetics. The meta-GGA functionals, SCAN and r$^2$SCAN, yield intermediate results, predicting formation energies at the VBM of around 5.1 eV and 5.5 eV, respectively, with the KO correction. A clear trend is observed where the formation energy systematically increases in the order PBE \textless SCAN \textless r$^2$SCAN \textless HSE06, suggesting that more sophisticated treatments of exchange and correlation stabilize the perfect crystal structure relative to the defected one. As the Fermi level moves towards the conduction band minimum (CBM), the formation energies for silicon vacancies decrease sharply across all functionals, indicating a strong sensitivity to the electronic environment and suggesting that Si vacancies become more energetically favorable under n-type conditions. The charge transition levels, which represent the Fermi energy values where different charge states become the most stable, are found at consistent positions relative to the VBM across the different functionals. \textcolor{blue}{Specifically, for Si vacancies (Fig. \ref{fig:formation}(c)), the (+1/0) transition level occurs at approximately 1.0-1.2 eV above the VBM for HSE06, SCAN, and $r^{2}$SCAN, while PBE predicts a notably lower value of 0.43 eV. This discrepancy likely arises from PBE’s underestimation of the band gap and its tendency to overdelocalize defect states, leading to shallower transition levels. The (0/-1) transition level is found at approximately 2.2-2.3 eV for HSE06, 1.37 eV for SCAN, 1.69 eV for $r^{2}$SCAN, and 0.84 eV for PBE, as seen in Fig. \ref{fig:formation}(c).} The slopes of the formation energy lines for the charged states ($+1$ and $-1$) are exactly $+1$ and $-1$, respectively, consistent with the theoretical formulation (Equation 1) and reflecting the physical process of charge transfer. These transition levels suggest that silicon vacancies can behave as both donors and acceptors within 1L-SiC. Figure \ref{fig:formation}(c) provides a bar chart summarizing these charge transition levels for Si vacancies across the four functionals.
	
	Carbon vacancies (Fig. \ref{fig:formation}(b)) consistently exhibit lower formation energies compared to silicon vacancies across all functionals and charge states shown. Using the KO correction (solid lines) at the VBM, the formation energies for C vacancies range from about 4.0 eV (PBE) to about 5.1 eV (HSE06). This lower energy cost compared to Si vacancies (e.g., approximately 5.1 eV for $V_{C}$ vs 6.0 eV for $V_{Si}$ using HSE06 with KO at the VBM) reflects the relative energetics of removing a C versus a Si atom. The dependence of the formation energy on the functional follows a similar pattern to that observed for Si vacancies, with PBE predicting the lowest energies and HSE06 the highest.The charge transition levels for C vacancies (Fig. 1(d)) show the (+1/0) transition occurring near 0.8-0.9 eV above the VBM, and the (0/-1) transitions occurring around 1.1 eV for PBE, SCAN, and  r$^2$SCAN, and around 1.6-1.7 eV for HSE06. However, an interesting difference is that the formation energy of carbon vacancies decreases less steeply with an increasing Fermi level compared to silicon vacancies. For example, using HSE06 (KO), the formation energy for C vacancies decreases by about 1.3 eV over the first eV increase in Fermi level, compared to a steeper ~1.8 eV decrease for Si vacancies over the same range. This suggests a reduced sensitivity to the electronic environment for C vacancies and implies they might be less effective as compensating defects in n-type conditions compared to Si vacancies. Figure \ref{fig:formation}(d) presents a bar chart illustrating these charge transition levels for C vacancies for each functional. 
	
	Comparing the charge correction schemes, the Kumagai-Oba (KO) method\cite{kumagai2014electrostatics} (solid lines) consistently yields slightly higher formation energies (by ~0.1-0.2 eV) than the Freysoldt-Neugebauer-Van de Walle (FNV) method\cite{freysoldt2009fully} (dashed lines) across most functionals and Fermi levels shown in the figure. This discrepancy stems from the KO method's more thorough treatment of the anisotropic dielectric screening inherent in 2D materials, utilizing the explicitly calculated in-plane ($\epsilon_{||} = 4.2$) and out-of-plane ($\epsilon_{\perp} = 1.8$) dielectric constants. The difference between the correction schemes underscores the importance of employing appropriate charge correction methods for accurate defect calculations in 2D systems. The thermodynamic stability window, indicating the range of Fermi levels where these vacancies can exist in thermal equilibrium, spans from the VBM up to approximately 1.6-1.7 eV (the(0/-1) transition level for HSE06) for C vacancies, and up to approximately 2.2-2.3 eV (the (0/-1) transition level for HSE06) for Si vacancies under the assumed carbon-rich conditions, when considering these $V^{0/-}$ transition levels as the upper limits within the displayed energy range of the figure. This range is constrained by the formation of competing defects and the material's intrinsic band gap, although the absolute values of the formation energies are dependent on the specific chemical potential references chosen for silicon and carbon.

	\subsection{Electronic Properties}
	
	The electronic band structure of pristine monolayer silicon carbide (1L-SiC) provides fundamental insights into its electronic properties and potential applications. Figure 2 presents the band structures calculated using PBE, SCAN, r$^2$SCAN, and HSE06 functionals. All four functionals consistently predict that 1L-SiC is an indirect bandgap semiconductor, with the valence band maximum (VBM) located at the $\Gamma$ point and the conduction band minimum (CBM) at the M-point of the Brillouin zone. The calculated HSE06 bandgap of 3.39 eV compares well with previous HSE06 calculations (3.35-3.58 eV)\cite{ha2023novel,mohseni2024vacancy} and is somewhat lower than GW predictions (3.90 eV).\cite{lu2012tuning} The indirect nature of the bandgap is a key feature of 1L-SiC, distinguishing it from other 2D materials like monolayer transition metal dichalcogenides, and has important implications for its optical and electronic properties. The indirect bandgap in 1L-SiC necessitates phonon assistance for optical transitions to conserve momentum. This three-particle process (electron, hole, and phonon) is less probable than the two-particle process in direct bandgap materials, leading to longer radiative lifetimes. The phonon involvement allows the system to remain in the excited state for a longer duration, as the electron must wait for an appropriate phonon to assist in the recombination process. This mechanism can be advantageous for applications requiring sustained excited states, such as in certain photocatalytic processes or in devices exploiting long-lived excitons. For pristine 1L-SiC, the spin-up (pink) and spin-down (blue) bands are perfectly degenerate across all functionals, indicating the absence of any intrinsic magnetic moment in the unperturbed 1L-SiC lattice. This spin degeneracy is expected for a non-magnetic semiconductor and serves as a baseline for understanding the spin-polarized states that emerge when defects are introduced. The band dispersion characteristics show remarkable qualitative similarity across all functionals, with some quantitative differences. The valence bands exhibit strong dispersion near the $\Gamma$-point, particularly evident in the topmost valence band. This strong dispersion indicates high hole mobility in the $\Gamma$-point region, suggesting that hole transport could play a significant role in the electronic properties of p-type or ambipolar 1L-SiC devices. In contrast, the conduction bands show noticeably less dispersion, especially near the CBM at the M-point. This flatter conduction band suggests lower electron mobility compared to hole mobility, an asymmetry that could have important implications for charge transport in 1L-SiC-based devices. The lower electron mobility might result in longer electron lifetimes, which could be advantageous for applications requiring sustained electron excitation, such as in photocatalysis or certain optoelectronic devices. Table 1 provides a comprehensive summary of key electronic properties calculated using each functional, offering valuable insights into these differences. The PBE functional, employing the GGA, predicts the smallest bandgap of 2.54 eV, in good agreement with previous studies.\cite{Bekaroglu2010,ha2023novel} This underestimation is a well-known limitation of GGA functionals, stemming from their approximate treatment of exchange-correlation effects and the lack of proper self-interaction correction.
	
	\begin{table*}[t]
		\centering
		\begin{tabular}{|l|c|c|c|c|}
			\hline
			\textbf{DFT} & \textbf{$E_{g}$ (eV)} & \textbf{WF (eV)} & \textbf{EA (eV)} & \textbf{IE (eV)} \\ \hline
			PBE & 2.54 & 4.85 & 2.46 & 5.10 \\ \hline
			SCAN & 2.81 & 4.99 & 2.36 & 5.29 \\ \hline
			r$^2$SCAN & 2.86 & 5.03 & 2.31 & 5.28 \\ \hline
			HSE06 & 3.39 & 4.79 & 2.47 & 5.03 \\ \hline
			
		\end{tabular}
		\caption{Properties of pristine SiC monolayer computed using DFT functionals.}
		\label{tab:sic_monolayer}
	\end{table*}
	Moving to more sophisticated functionals, we observe a systematic increase in the predicted bandgap. The SCAN functional, a meta-GGA approach, yields a larger bandgap of 2.81 eV, while its refined version, r$^2$SCAN, predicts a slightly higher value of 2.86 eV. This trend reflects the improved treatment of exchange-correlation effects in meta-GGA functionals, which incorporate the kinetic energy density to better account for electron localization effects. The HSE06 hybrid functional, which includes a fraction of exact exchange, predicts the largest bandgap of 3.39 eV. This value is expected to be the most accurate among the functionals considered, as hybrid functionals are known to provide more reliable bandgap predictions for semiconductors due to their improved treatment of the self-interaction error. Beyond the bandgap, Table 1 reveals interesting trends in other electronic properties. The work function, a critical parameter for understanding electron emission and interface properties, shows less variation across functionals compared to the bandgap. PBE predicts a work function of 4.85 eV, while SCAN and r$^2$SCAN predict slightly higher values of 4.99 eV and 5.03 eV, respectively. Interestingly, HSE06 predicts a slightly lower work function of 4.79 eV compared to the meta-GGA functionals. The electron affinity follows a non-monotonic trend across functionals. PBE and HSE06 predict similar values (2.46 eV and 2.47 eV, respectively), while SCAN and r$^2$SCAN predict lower values (2.36 eV and 2.31 eV). This suggests that meta-GGA functionals predict a lower tendency for 1L-SiC to accept electrons compared to both PBE and HSE06. The ionization energy demonstrates a different pattern, with SCAN and r$^2$SCAN predicting the highest values (5.29 eV and 5.28 eV), while PBE and HSE06 predict lower values (5.10 eV and 5.03 eV). These differences in electron affinity and ionization energy could significantly influence predictions of the material's chemical reactivity, charge transfer processes, and behavior in electronic devices. The introduction of silicon vacancies in 1L-SiC induces profound changes in its electronic structure, as illustrated in Figure 3 and supported by density of states (DOS) analysis in the Supplementary Information (Figure S4-S5). For the neutral vacancy V$^0_\text{Si}$ (Figure 3a), our calculations reveal significant functional dependence in the predicted electronic structure. Using the PBE functional, we observe a metallic state with bands crossing the Fermi level. These bands, derived primarily from dangling bond states of neighboring carbon atoms, exhibit minimal dispersion ($\approx$  0.2 eV bandwidth), indicating strong spatial localization of the defect states. This localization is evidenced in the DOS by sharp peaks with full width at half maximum of approximately 0.3 eV.
	\begin{figure}[H]
		\centering
		\includegraphics[width=0.9\textwidth]{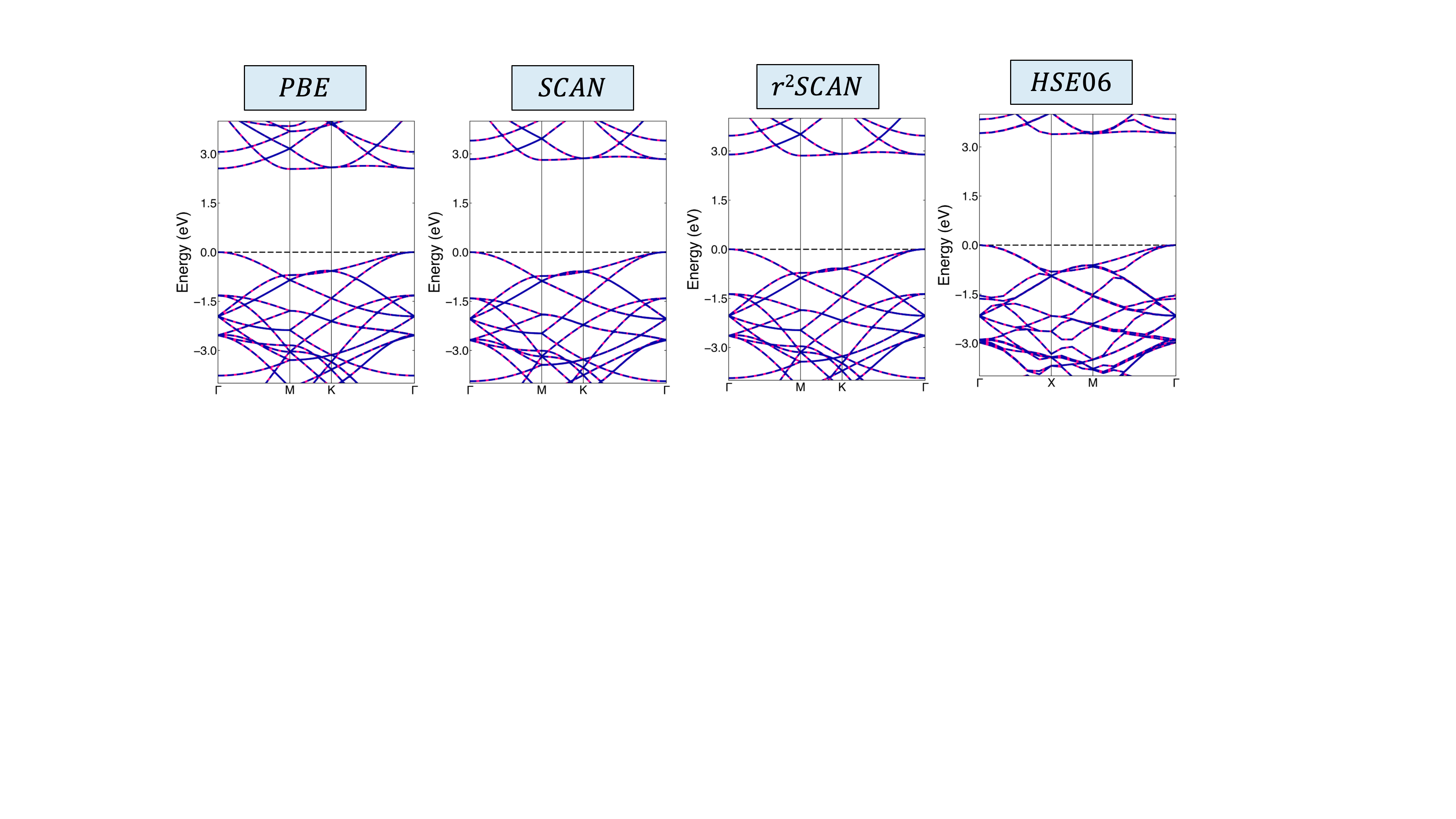}
		\caption{Electronic band structures of pristine SiC monolayer calculated using PBE, SCAN, r$^2$SCAN, and HSE06 functionals (from left to right). The Fermi level is set to zero. Red dashed lines indicate the indirect bandgap from $\Gamma$ to M. The progressive increase in bandgap from PBE (2.54 eV) to HSE06 (3.39 eV) demonstrates the impact of improved exchange-correlation treatment.}
		\label{fig:pristine}
	\end{figure}
	
	In contrast, SCAN and r$^2$SCAN reveal the emergence of spin polarization, with differing band structures for spin-up and spin-down channels. Both meta-GGA functionals predict small gaps in the spin-up channel, while maintaining partially occupied states in the spin-down channel. This spin polarization, absent in PBE results, demonstrates the importance of improved exchange-correlation treatment for capturing the magnetic properties of defect states. The HSE06 functional reveals the most dramatic electronic structure, with the spin-up bands exhibiting a gap of approximately 2.0 eV between the VBM at $\Gamma$ and the defect states, while the spin-down bands show metallic character with states crossing the Fermi level. To verify convergence of these electronic states, we performed multiple calculations starting from different initial densities and with various initial magnetic moments, confirming that the HSE06 result represents the true ground state. The DOS analysis reveals a clear spin splitting of 0.7 eV between majority and minority spin channels in HSE06 calculations, providing a quantitative measure of the magnetic moment. This substantial spin splitting, progressively developing from PBE to HSE06, underscores the critical role of exact exchange in correctly describing the spin physics of localized defect states. For the negatively charged silicon vacancy (V$^{-1}_\text{Si}$), shown in Figure 3b, we observe even more pronounced electronic structure changes across functionals. With PBE, the additional electron leads to increased occupation of defect states without significant modification of their dispersion. However, SCAN and r$^2$SCAN both predict the opening of distinct gaps in the spin-up and spin-down channels, with clear spin polarization evident in the band structure. The HSE06 results for V$^{-1}_\text{Si}$ exhibit the most dramatic spin polarization. A large indirect gap of 2.2 eV appears in the spin-up channel, while the spin-down channel shows a smaller gap of 0.9 eV. This spin splitting of 1.3 eV is the largest observed among all vacancy configurations studied, indicating that the additional electron enhances the magnetic character of the defect. The DOS reveals sharply defined peaks split by 1.3 eV between spin channels, with minimal broadening ($\approx$ 0.2 eV width), consistent with the highly localized nature of these states. This strong spin-dependence indicates that the additional electron preferentially occupies one spin channel, maximizing the exchange energy and resulting in a high-spin configuration. For the positively charged vacancy V$^{+1}_\text{Si}$ (Figure 3c), we observe yet another distinct electronic structure. PBE calculations predict reduced spin polarization compared to the neutral case, with the removal of an electron leading to more symmetric occupation of spin channels. SCAN and r$^2$SCAN maintain moderate spin polarization, with spin-dependent gaps of varying magnitude. The HSE06 functional predicts a wide gap of 3.0 eV for spin-up states and metallic behavior for spin-down states, creating a highly asymmetric electronic structure. This stark asymmetry between spin channels implies a complex redistribution of electrons upon positive charging, possibly involving significant lattice relaxation and rehybridization of surrounding atoms. Orbital analysis of the defect states reveals that silicon vacancy states are dominated by p-orbital contributions from the surrounding carbon atoms, with p:s peak intensity ratios exceeding 3:1 in the defect state region. This strong p-orbital character persists across all charge states, though the relative weights shift with charging: the p-orbital contribution increases to nearly 85\% in V$^{-1}_\text{Si}$ but decreases to approximately 70\% in V$^{+1}_\text{Si}$, reflecting charge-state-dependent changes in the local electronic structure.
	
	Carbon vacancies demonstrate distinctive electronic characteristics that significantly differ from silicon vacancies, reflecting the fundamental differences in C-Si and Si-Si bonding in the SiC lattice. Figure 4 presents the band structures for carbon vacancies in different charge states, calculated with our four functionals. For the neutral carbon vacancy (V$^0_\text{C}$), PBE calculations predict a small spin splitting, with slightly different band structures for the two spin channels. SCAN and r$^2$SCAN enhance this spin polarization, with more pronounced differences between spin-up and spin-down bands, particularly near the Fermi level.
	
	\begin{figure}[H]
		\centering
		\includegraphics[width=0.9\textwidth]{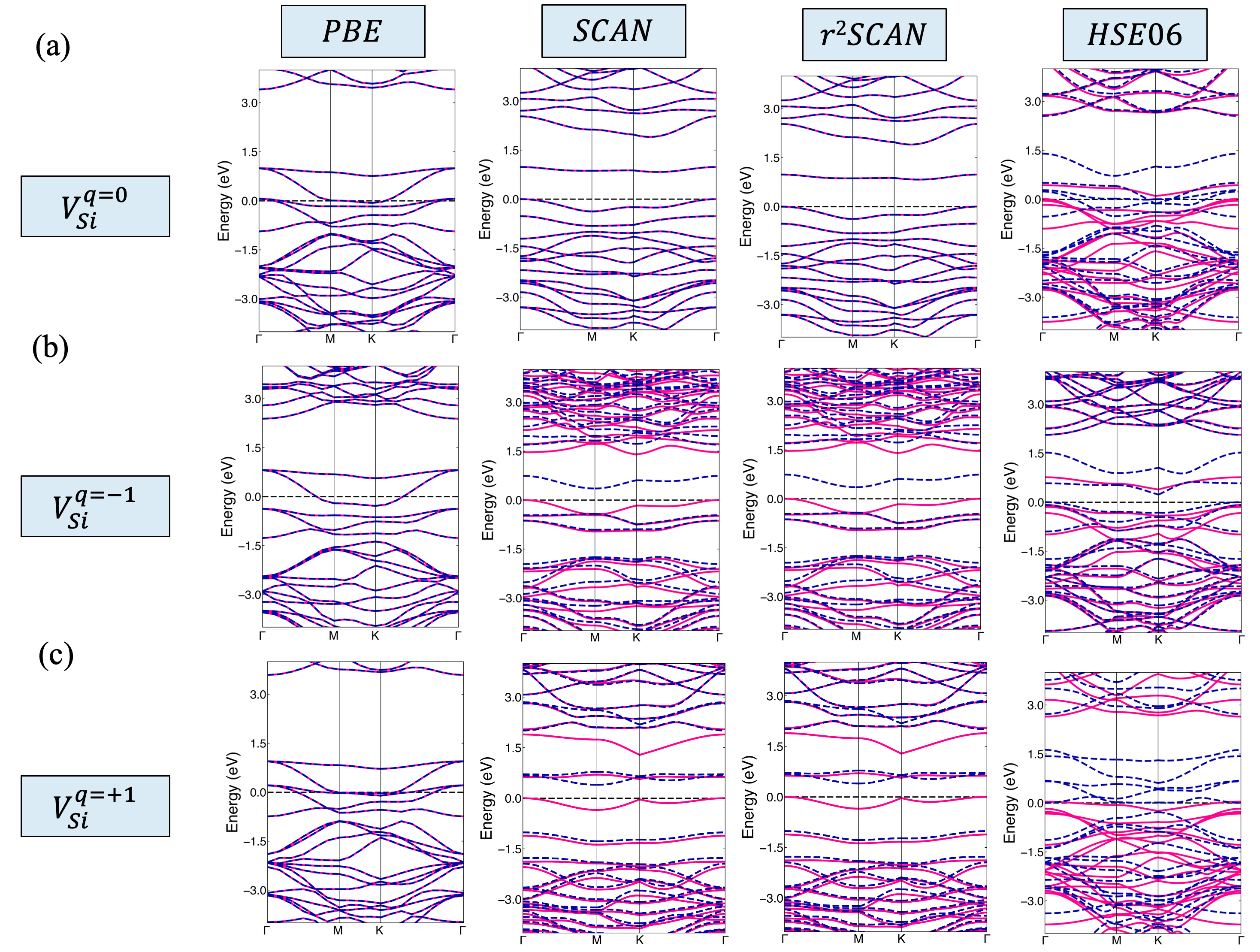}
		\caption{Electronic band structure of Si vacancies in a 3×3 monolayer SiC computed using PBE, SCAN, r$^2$SCAN and HSE06 functionals (with D3 dispersion correction) for three charge states: (a) V$^0_\text{Si}$, (b) V$^{-1}_\text{Si}$, and (c) V$^{+1}_\text{Si}$. Pink and blue lines represent spin-up and spin-down bands, respectively. The Fermi level is set to zero. Note the progressive increase in spin splitting from PBE to HSE06, with the latter showing the most pronounced spin polarization, particularly for V$^{-1}_\text{Si}$.}
		\label{fig:si_vacancy}
	\end{figure}
	HSE06 calculations reveal a unique electronic structure with a moderate spin splitting of 0.3 eV, substantially smaller than the 0.7 eV observed for V$^0_\text{Si}$. This is evidenced by both the band structure gaps (1.5 eV for spin-up and 1.2 eV for spin-down channels) and the corresponding DOS peaks separated by 0.3 eV near the Fermi level. The defect states show broader dispersion of approximately 0.5 eV across the Brillouin zone, particularly notable along the $\Gamma$-K direction, compared to the much narrower 0.2 eV dispersion observed for silicon vacancies. This broader bandwidth indicates stronger hybridization with neighboring atoms and a more delocalized character of the defect states. DOS analysis reveals that carbon vacancy states have significant contributions from both s and p orbitals of the surrounding silicon atoms, with p-orbital peaks showing approximately 50\% greater intensity than s-orbital peaks in the energy range of -1 to 1 eV around the Fermi level. This mixed orbital character explains the enhanced band dispersion and suggests stronger hybridization with neighboring atoms compared to the predominantly p-orbital character of silicon vacancy states. The negatively charged carbon vacancy (V$^{-1}_\text{C}$) exhibits a complex electronic structure that varies significantly across functionals. PBE predicts metallic behavior with multiple band crossings at the Fermi level, while SCAN and r$^2$SCAN reveal the emergence of small gaps (0.4-0.6 eV) in both spin channels.
	\begin{figure}[H]
		\centering
		\includegraphics[width=0.9\textwidth]{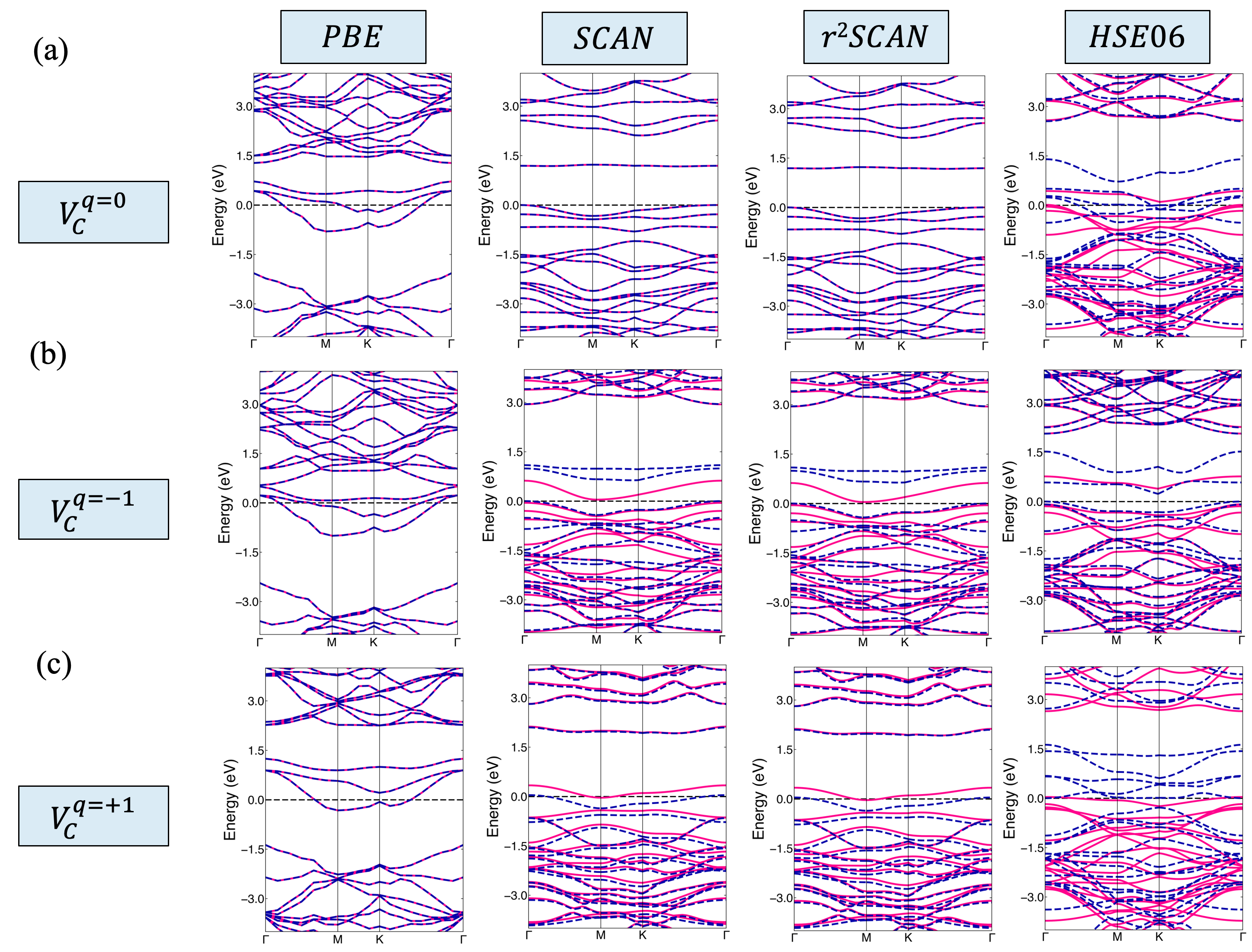}
		\caption{Electronic band structure of C vacancies in a 3×3 monolayer SiC computed using PBE, SCAN, r$^2$SCAN and HSE06 functionals (with D3 dispersion correction) for three charge states: (a) V$^0_\text{C}$, (b) V$^{-1}_\text{C}$, and (c) V$^{+1}_\text{C}$. Pink and blue lines represent spin-up and spin-down bands, respectively. The Fermi level is set to zero. Carbon vacancy states show broader dispersion than silicon vacancies, indicating more delocalized defect states with weaker spin polarization.}
		\label{fig:c_vacancy}
	\end{figure}
	
	The HSE06 calculations demonstrate spin-split gaps of 0.8 eV and 0.5 eV for spin-up and spin-down channels respectively, notably smaller than the corresponding gaps in V$^{-1}_\text{Si}$ (2.2 eV and 0.9 eV). The DOS shows three distinct peaks at -0.8 eV, -0.2 eV, and +0.3 eV relative to the Fermi level, corresponding to specific transitions in the band structure. For the positively charged state (V$^{+1}_\text{C}$), HSE06 calculations reveal a direct gap of 0.6 eV at the K point for the spin-up channel, while maintaining metallic character in the spin-down channel. The DOS exhibits a sharp peak at +0.4 eV above the Fermi level in the spin-down channel, with an integrated intensity approximately twice that of the neutral state peaks, indicating enhanced state density available for optical transitions. The progression of electronic structure across charge states shows systematic trends in the DOS peak positions: the main defect state peaks shift from -0.3 eV in V$^{-1}_\text{C}$ to 0.0 eV in V$^0_\text{C}$ to +0.4 eV in V$^{+1}_\text{C}$ relative to the Fermi level, demonstrating clear charge state-dependent electronic structure evolution. This detailed analysis of band structures and DOS reveals that carbon vacancies maintain moderate spin polarization (0.3-0.8 eV splittings) across charge states, but with consistently smaller magnitudes than silicon vacancies (0.7-1.3 eV splittings). The broader bandwidth of carbon vacancy states (0.5 eV vs 0.2 eV) and mixed s-p orbital character indicate fundamentally different bonding characteristics, leading to more delocalized defect states that could influence their optical and magnetic properties.
	
	\subsection{Migration of Vacancies}
	Figure 5 presents our comprehensive investigation of vacancy migration mechanisms in monolayer SiC using nudged elastic band (NEB) calculations with the HSE06 functional including D3 dispersion correction. Figures 5(a) and 5(b) illustrate the top view of the monolayer SiC structure, showing the distinct migration pathways for Si and C vacancies, respectively. For both vacancy types, we considered three mechanistically different pathways: Path A (green) represents migration to the diagonal position of the same atom type within a single hexagon, Path B (pink) follows migration to the same atom type position diagonally crossing two hexagons, and Path C (blue) describes migration to the nearest neighbor of the same atom type. The energy profiles along these migration pathways for Si vacancies (Fig. 5c) and C vacancies (Fig. 5d) reveal significant differences in migration barriers. For Si vacancies, our calculations show that Path C exhibits the lowest energy barrier of 0.8 eV, making it the most favorable migration route. Path A presents an intermediate barrier of 1.25 eV, while Path B shows the highest barrier of 1.75 eV, making it the least likely route for Si vacancy migration. These values represent significant revisions to previously reported barriers, reflecting our more accurate treatment of transition state geometries and careful convergence of the NEB calculations. For C vacancies, the calculated migration barriers are: Path C presents the lowest energy barrier at approximately 1.0 eV, followed by Path A with approximately 1.2 eV, and Path B with approximately 1.4 eV. The consistently higher barriers for C vacancies compared to Si vacancies suggest that C vacancies are fundamentally less mobile in the 1L-SiC lattice. Examination of the transition state (TS) configurations provides crucial insights into the atomic-scale mechanisms of vacancy migration. For Si vacancies following Path C, the TS involves a displacement of the migrating silicon atom towards the vacancy site by approximately 0.8 $\AA$, accompanied by a modest out-of-plane distortion of 0.2 $\AA$. The small magnitude of these distortions contributes to the relatively low barrier height. The transition state exhibits C$_{2v}$ symmetry, maintaining two mirror planes perpendicular to the SiC sheet. In contrast to Path C, the transition state for Path B shows substantially larger structural distortions. For Si vacancies, the TS structure exhibits significant in-plane lattice distortion, with neighboring atoms displaced by up to 0.5 $\AA$, and more pronounced out-of-plane buckling reaching 0.4 $\AA$. This extensive reorganization explains the higher barrier of 1.75 eV. The transition state loses all mirror symmetry elements, adopting a C$_1$ point group that reflects the complex atomic rearrangements required for migration across two hexagons. The surrounding atoms show collective distortions extending to second and third nearest neighbors, indicating the long-range nature of the structural response. Path A presents an intermediate case, with the transition state showing moderate distortions. For Si vacancies, the migrating atom undergoes a displacement of approximately 0.6 $\AA$ in-plane and 0.3 $\AA$ out-of-plane, maintaining a single mirror plane (C$_s$ symmetry). The resulting barrier of 1.25 eV reflects this intermediate level of structural reorganization. Notably, the transition state retains partial symmetry, unlike the lowest-symmetry configuration found in Path B. For carbon vacancies, the transition states show systematically larger distortions across all pathways, correlating with their higher migration barriers. The Path C transition state for C vacancies involves a 1.0 $\AA$ displacement of the migrating carbon atom, with out-of-plane distortion reaching 0.3 $\AA$. This increased distortion compared to Si vacancies (0.8 $\AA$ and 0.2 $\AA$ respectively) helps explain the barrier of approximately 1.0 eV. The stronger covalent bonding character of carbon atoms requires more energy for bond breaking and reformation during migration. The Path B transition state for C vacancies shows the most extreme structural changes, with in-plane distortions reaching 0.7 $\AA$ and out-of-plane buckling of 0.6 $\AA$. This extensive reconstruction results in the barrier of approximately 1.4 eV. The large distortions reflect the significant energy cost of breaking multiple C-Si bonds simultaneously during migration across two hexagons. The transition state geometry shows complex rehybridization of carbon orbitals, evidenced by significant changes in C-Si bond angles from the ideal 120$^{o}$ to values ranging from 105$^{o}$ to 135$^{o}$. A comparative analysis of Si and C vacancy migration mechanisms reveals fundamental differences in their behavior. Silicon vacancies generally maintain more rigid local structures during migration, with smaller distortions of the surrounding lattice. This rigidity can be attributed to the more ionic character of Si-C bonds and the larger size of the silicon atom, which results in smaller relative displacements during migration. The energy cost for breaking and reforming Si-C bonds is also lower, contributing to the generally smaller barriers for Si vacancy migration.
	\begin{figure}[H]
		\centering
		\includegraphics[width=0.9\textwidth]{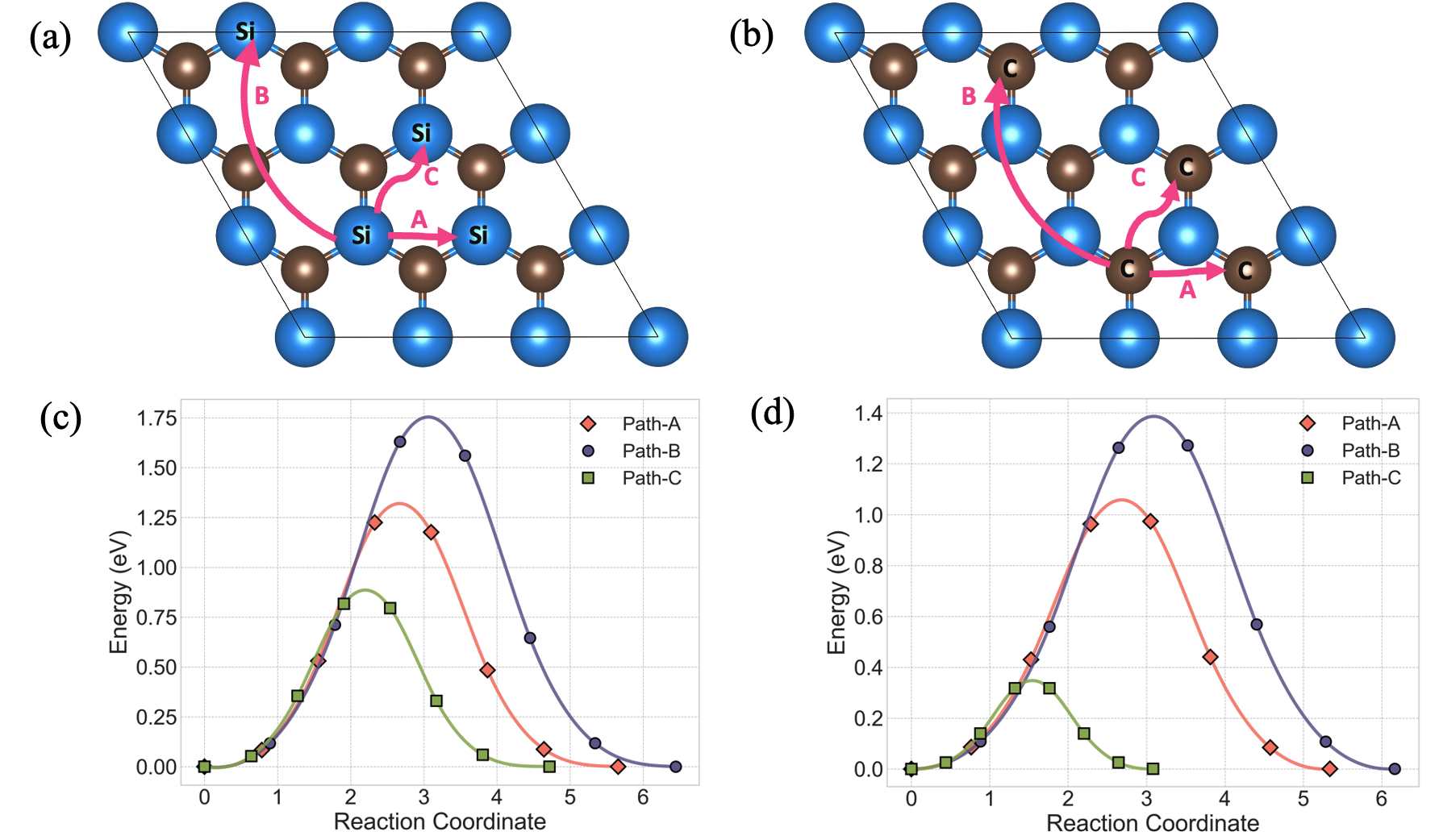}
		\caption{Nudged elastic band (NEB) calculations of vacancy migration pathways in 1L-SiC. (a) Si vacancy and (b) C vacancy migration paths. Path A (green): migration within single hexagon; Path B (pink): across two hexagons; Path C (blue): to nearest neighbor. Energy profiles for (c) Si vacancy and (d) C vacancy migration shown relative to initial state energy. Si vacancies exhibit barriers ranging from 0.8 eV to 1.75 eV, while C vacancies have barriers ranging from approximately 1.0 eV to 1.4 eV. This generally indicates higher mobility for Si vacancies, particularly as the lowest energy path for Si (0.8 eV) is lower than for C (1.0 eV).}
		\label{fig:migration}
	\end{figure}
	
	Carbon vacancies, by contrast, induce more pronounced structural distortions during migration due to the stronger covalent character of their bonds and the smaller atomic radius of carbon. The migration process for carbon vacancies involves significant rehybridization of orbitals and more substantial rearrangement of the surrounding lattice, resulting in consistently higher energy barriers. These differences in migration barriers have important implications for defect dynamics in 1L-SiC. he significantly lower barrier for Si vacancy migration (0.8 eV via Path C) compared to C vacancies (approximately 1.0 eV via Path C) suggests that Si vacancies will be more mobile at moderate temperatures. Using transition state theory with a typical attempt frequency of 10$^{13}$ s$^{-1}$, we can estimate diffusion rates at various temperatures. At room temperature (300 K), Si vacancies (with $E_{b}=$0.8 eV) would have a hopping rate of approximately 2.5 × 10$^{-1}$ s$^{-1}$. If the barrier for C vacancies (Path C) is approximately 1.0 eV, their hopping rate would be around 1.5 × 10$^{-8}$ s$^{-1}$. At higher temperatures such as 600 K, the difference becomes less pronounced but still significant: Si vacancies would have a hopping rate on the order of 10$^5$- 10$^6$ s$^{-1}$, while C vacancies would diffuse at about 3.6 × 10$^4$ s$^{-1}$. These differences in mobility could lead to preferential aggregation of Si vacancies or formation of vacancy complexes under thermal annealing conditions, potentially influencing the evolution of defect structures in 2D SiC. The higher mobility of Si vacancies may also explain the experimental observation of divacancy complexes (V$_\text{Si}$-V$_\text{C}$) in SiC,\cite{Pizzochero2019} as mobile Si vacancies could migrate towards relatively immobile C vacancies and form stable complexes.
	
	\subsection{Optical Properties}
	
	The introduction of vacancies in monolayer SiC (1L-SiC) induces profound changes in its optical properties. Figure 6 presents our analysis of the dielectric response and absorption characteristics, computed using density functional perturbation theory (DFPT) within the PBE framework. We calculated both the real ($\varepsilon_1$) and imaginary ($\varepsilon_2$) parts of the dielectric function (Figures 6a and 6c), along with the corresponding optical absorptivity (Figures 6b and 6d). For pristine 1L-SiC, $\varepsilon_2$ shows a prominent peak at approximately 4 eV, corresponding to fundamental optical transitions across the bandgap. The real part of the dielectric function ($\varepsilon_1$) exhibits characteristic oscillator behavior, with its magnitude directly influencing the material's refractive index. For the neutral silicon vacancy (V$^0_\text{Si}$), we observe a distinct peak at 0.1 eV in both $\varepsilon_1$ and $\varepsilon_2$, with $\varepsilon_2$ reaching approximately 2-3. This peak position aligns with transitions between the defect states identified in our electronic structure calculations. The absorption coefficient of V$^{0}_\text{Si}$ shows enhancement of approximately 3-4\% around 0.07 eV. The negatively charged silicon vacancy (V$^{-1}_\text{Si}$) produces a broad response below 0.1 eV in both $\varepsilon_1$ and $\varepsilon_2$, with multiple peaks of magnitude around 5. This electronic structure results in enhanced absorption in the far-infrared region, showing peaks of 3-4\% at energies below 0.1 eV. These features correlate directly with the multiple transitions available between spin-split states identified in our electronic structure analysis. For the positively charged state (V$^{+1}_\text{Si}$), we find the strongest dielectric response, with $\varepsilon_2$ showing peaks reaching 20 at energies below 0.1 eV. This translates to the highest absorption among all vacancy states, with enhancement up to 12\% in the far-infrared region. The absorption profile shows multiple distinct bands between 0.1-0.3 eV, reflecting transitions between the asymmetric spin channels observed in the electronic structure. These multiple absorption features correspond to transitions between the well-defined defect levels in the band structure of V$^{+1}_\text{Si}$ shown in Figure 3c. Carbon vacancies demonstrate distinctly different optical responses compared to silicon vacancies, reflecting their more delocalized electronic structure. For the neutral carbon vacancy (V$^0_\text{C}$), we observe moderate peaks in $\varepsilon_2$ ($\approx$ 4), resulting in absorption coefficients of 3-4\%. These relatively weak oscillator strengths correlate with the broader dispersion of carbon vacancy states identified in our band structure analysis. The absorption features appear more distributed across the far-infrared region, lacking the sharp peaks characteristic of silicon vacancies. The negatively charged carbon vacancy (V$^{-1}_\text{C}$) shows enhanced absorption (5-6\%) compared to the neutral state, arising from additional transitions enabled by the extra electron. The dielectric response exhibits three distinct features corresponding to the DOS peaks identified at -0.8 eV, -0.2 eV, and +0.3 eV relative to the Fermi level. These transitions produce a more complex absorption spectrum than V$^{-1}_\text{Si}$, with multiple overlapping bands below 0.2 eV, reflecting the more delocalized nature of the defect states. The positively charged state (V$^{+1}_\text{C}$) demonstrates the strongest response among carbon vacancy configurations, with $\varepsilon_2$ reaching approximately 20 below 0.05 eV. This enhancement results in absorption peaks exceeding 20\%, making it particularly promising for far-infrared applications. Unlike the discrete absorption bands seen in V$^{+1}_\text{Si}$, V$^{+1}_\text{C}$ exhibits a more continuous absorption profile extending up to 0.4 eV, reflecting its more delocalized electronic structure.
	\begin{figure}[H]
		\centering
		\includegraphics[width=0.9\textwidth]{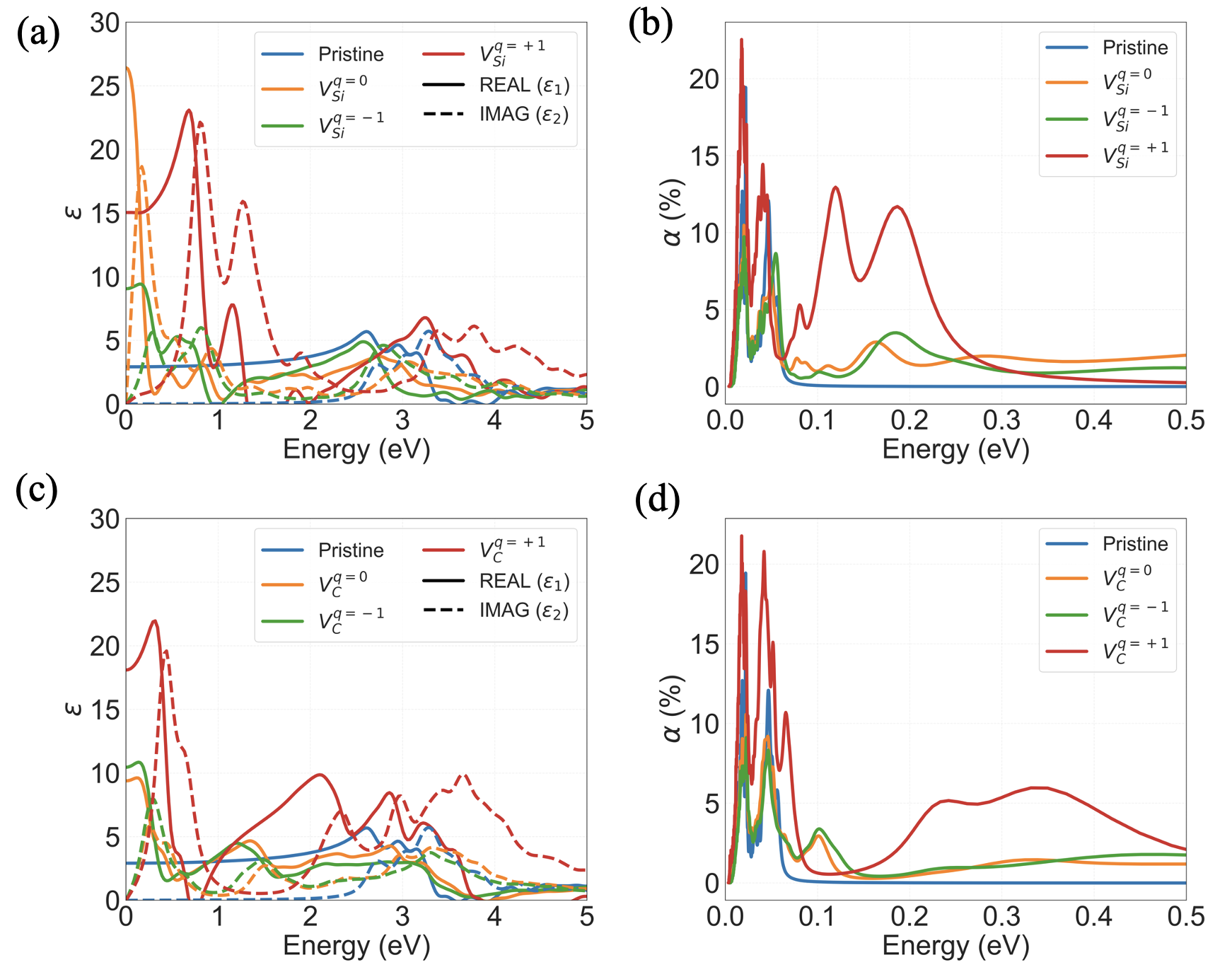}
		\caption{Optical properties of vacancy defects in 1L-SiC computed using PBE-DFPT: (a,c) Real ($\varepsilon_1$, solid lines) and imaginary ($\varepsilon_2$, dashed lines) parts of the dielectric function for Si and C vacancies respectively in different charge states. The pristine case is shown in blue for reference. (b,d) Corresponding absorption coefficients ($\alpha$) shown as percentage change relative to pristine SiC. Positively charged vacancies exhibit the strongest optical response, with V$^{+1}_\text{C}$ showing absorption peaks reaching approximately 21\%, and V$^{+1}_\text{Si}$ showing peaks up to approximately 12\%, in the far-infrared region (0-0.5 eV).}
		\label{fig:optical}
	\end{figure}
	
	Local field effects were fully incorporated in our calculations through the inclusion of off-diagonal elements of the dielectric matrix up to 100 eV in the plane-wave basis. The frequency-dependent dielectric response was computed up to 5 eV using a dense frequency grid of 0.02 eV to accurately capture fine spectral features. Temperature effects were included through the Gaussian broadening of 0.1 eV, chosen based on typical experimental linewidths in similar 2D materials. The contrasting optical behavior of Si and C vacancies can be understood through their distinct electronic structures. Silicon vacancies produce highly localized states leading to sharp optical transitions, while carbon vacancies generate more dispersed states resulting in broader absorption features. This difference is particularly evident in the positively charged states, where V$^{+1}_\text{Si}$ shows discrete absorption bands between 0.1-0.3 eV, while V$^{+1}_\text{C}$ exhibits a continuous absorption profile up to 0.4 eV. Our predictions are supported by recent experimental demonstrations of substrate-dependent optical properties in epitaxially grown 2D SiC.\cite{polley2023bottom} The strong charge-state dependence of optical properties aligns with theoretical predictions regarding excitonic effects in 2D SiC, particularly the enhancement of absorption strength through charge state modification. The magnitude of enhancement we predict (up to 21\% for V$^{+1}_\text{C}$ and around 12\% for V$^{+1}_\text{Si}$) exceeds typical values in other 2D materials, suggesting unique opportunities for optoelectronic applications in the far-infrared regime. The optical signatures we have identified could serve as spectroscopic fingerprints for different vacancy types and charge states, potentially enabling the identification and characterization of defects in experimental samples. Additionally, the strong charge-state dependence of optical properties suggests possibilities for electrostatic tuning of optical response, where the application of a gate voltage could modulate the charge state of defects and consequently their optical absorption properties. The far-infrared response of vacancy defects in 2D SiC could be particularly valuable for applications in gas sensing, thermal imaging, and infrared communication. The distinct absorption features in the 0.1-0.5 eV range (approximately 2.5-12 $\mu$m wavelength) coincide with important atmospheric transmission windows and molecular fingerprint regions, making these defects potentially useful for selective sensing applications.
	
	\section{Conclusions}
	This comprehensive study of silicon and carbon vacancies in monolayer SiC (1L-SiC) using density functional theory has provided crucial insights into the electronic, structural, and optical properties of these defects. Our investigation, employing a range of exchange-correlation functionals (PBE, SCAN, r$^2$SCAN, and HSE06) and charge correction schemes (FNV and KO), demonstrates the critical importance of methodology in accurately predicting defect properties in 2D materials. The formation energies of silicon and carbon vacancies reveal significant variations across functionals, with HSE06 consistently predicting the highest values (7.5-8.5 eV for Si vacancies and 6.0-7.0 eV for C vacancies at the VBM), highlighting the substantial impact of exact exchange inclusion on defect energetics. The relative stability of carbon vacancies over silicon vacancies by 1.0-1.5 eV persists across all functionals, providing a robust prediction independent of methodology. The charge transition levels, while differing quantitatively across functionals, maintain consistent relative positions, with the +1/0 and 0/-1 transitions occurring at approximately 1.0 eV and 2.2 eV above the VBM, respectively, for both vacancy types. The electronic structure analysis demonstrates distinct characteristics between silicon and carbon vacancies. Silicon vacancies introduce highly localized defect states, with HSE06 calculations revealing strong spin polarization effects and a gap of approximately 2 eV for spin-up states. In contrast, carbon vacancies exhibit more dispersed states with broader bandwidth ($\approx$ 0.5 eV), suggesting stronger orbital overlap with neighboring atoms. This fundamental difference in localization behavior has important implications for the potential applications of these defects in quantum technologies. The highly localized nature of silicon vacancy states makes them promising candidates for quantum bit (qubit) applications, potentially offering longer coherence times due to their reduced interaction with the environment, while carbon vacancies might be more suitable for applications requiring enhanced coupling to external fields.
	The progression of electronic structure with changing charge states follows systematic trends, with HSE06 calculations revealing significant spin polarization across all charge states for both vacancy types. The largest spin splitting of 1.3 eV is observed for V$^{-1}_\text{Si}$, suggesting that this configuration could exhibit the strongest magnetic properties among the studied systems. The orbital character analysis reveals predominantly p-orbital contributions for silicon vacancy states, while carbon vacancies show mixed s-p character, further explaining their different electronic properties. Our migration pathway analysis reveals significantly lower barriers for silicon vacancies (0.8 eV for the most favorable path) compared to carbon vacancies (1.2 eV), indicating higher mobility of silicon vacancies at moderate temperatures. This finding has important implications for defect dynamics and the potential formation of vacancy complexes in 1L-SiC, particularly under thermal annealing conditions where Si vacancy diffusion would dominate. The detailed transition state analysis provides atomic-level insights into the migration mechanisms, revealing that the stronger covalent bonding character of carbon atoms leads to more pronounced structural distortions during migration, explaining their higher barriers. The optical property calculations reveal significant modifications to the dielectric response and absorption characteristics, particularly in the far-infrared region (0-0.5 eV). Positively charged vacancies demonstrate the strongest optical response, with absorption coefficients reaching 22\% enhancement relative to pristine SiC. The distinct behavior of silicon vacancies (showing sharp, discrete absorption features) and carbon vacancies (exhibiting broader absorption profiles) provides complementary functionalities for potential optoelectronic applications. These optical signatures could serve as spectroscopic fingerprints for identifying and characterizing defects in experimental samples. \textcolor{blue}{The choice of computational methodology proves crucial for accurate predictions. While HSE06 provides the most reliable results for electronic structure and energetics among the functionals used, even more accurate methods, such as GW or diffusion Monte Carlo, exist but are computationally prohibitive for the periodic systems and large-scale defect calculations in this study \cite{lu2012tuning}. For molecular systems, coupled cluster methods like CCSD(T) serve as a gold standard, but their application to periodic crystals like 2D SiC is impractical. The computational efficiency of meta-GGA functionals like SCAN and $r^{2}$SCAN makes them attractive for initial screening studies.} The PBE-DFPT approach demonstrates reliability for optical property calculations while maintaining computational efficiency. The importance of appropriate charge correction schemes is underscored by the significant differences (up to 0.5 eV) between the FNV and KO methods, with the latter's treatment of anisotropic dielectric screening being particularly important for 2D materials. These findings open several promising directions for future research. The strong charge-state dependence of optical properties suggests possibilities for electrostatically tunable optical devices operating in the far-infrared region. The distinct characteristics of silicon and carbon vacancies could be exploited for dual-functionality devices combining sharp, state-specific optical transitions with broader absorption bands. The significant difference in mobility between Si and C vacancies suggests potential routes for controlled formation of vacancy complexes through selective thermal annealing. Future work should focus on exploring the interplay between different types of defects, investigating strain effects on defect properties, and experimental validation of these theoretical predictions. Studies of defect complexes, particularly the divacancy (V$_\text{Si}$-V$_\text{C}$) and transition metal dopant-vacancy pairs, could reveal novel quantum functionalities. Additionally, exploring the impact of external fields (electric, magnetic, strain) on defect properties could provide routes for dynamic control of quantum states. This work establishes a comprehensive framework for investigating defects in 2D materials and demonstrates how careful selection of computational methods can provide detailed insights into defect properties. The complementary functionalities of silicon and carbon vacancies revealed in this study suggest exciting possibilities for defect engineering in next-generation quantum and optoelectronic devices based on 2D SiC.
	
	\begin{acknowledgement}
		The author acknowledges the computational resources provided by the University of Delaware High Performance Computing facility. Special thanks to colleagues at the Delaware Energy Institute for valuable discussions.
	\end{acknowledgement}
	
	\begin{suppinfo}
		Supplementary information containing \textcolor{blue}{convergence test results and}detailed analysis of density of states (DOS) for all vacancy configurations, additional band structure plots, and transition state geometries for vacancy migration is available free of charge via the Internet at http://pubs.acs.org.
	\end{suppinfo}
	
	\section{Author Contributions}
	A. Patra conceived the study, performed calculations, and wrote the manuscript.
	
	\section{Conflicts of Interest}
	The author declares no competing financial interest.
	
	\bibliography{CompMatSci_ref.bib}

	\section{Highlights}
	\begin{itemize}
		\item Silicon and carbon vacancies in 2D SiC show distinct quantum properties
		\item Si vacancies exhibit highly localized states with strong spin polarization
		\item C vacancies demonstrate more dispersed states with weaker magnetic response
		\item Positively charged vacancies show strongest optical absorption up to 22\%
		\item HSE06 functional critical for accurate defect property predictions
	\end{itemize}
	
	\begin{figure}
		\centering
		\includegraphics[width=1.0\linewidth]{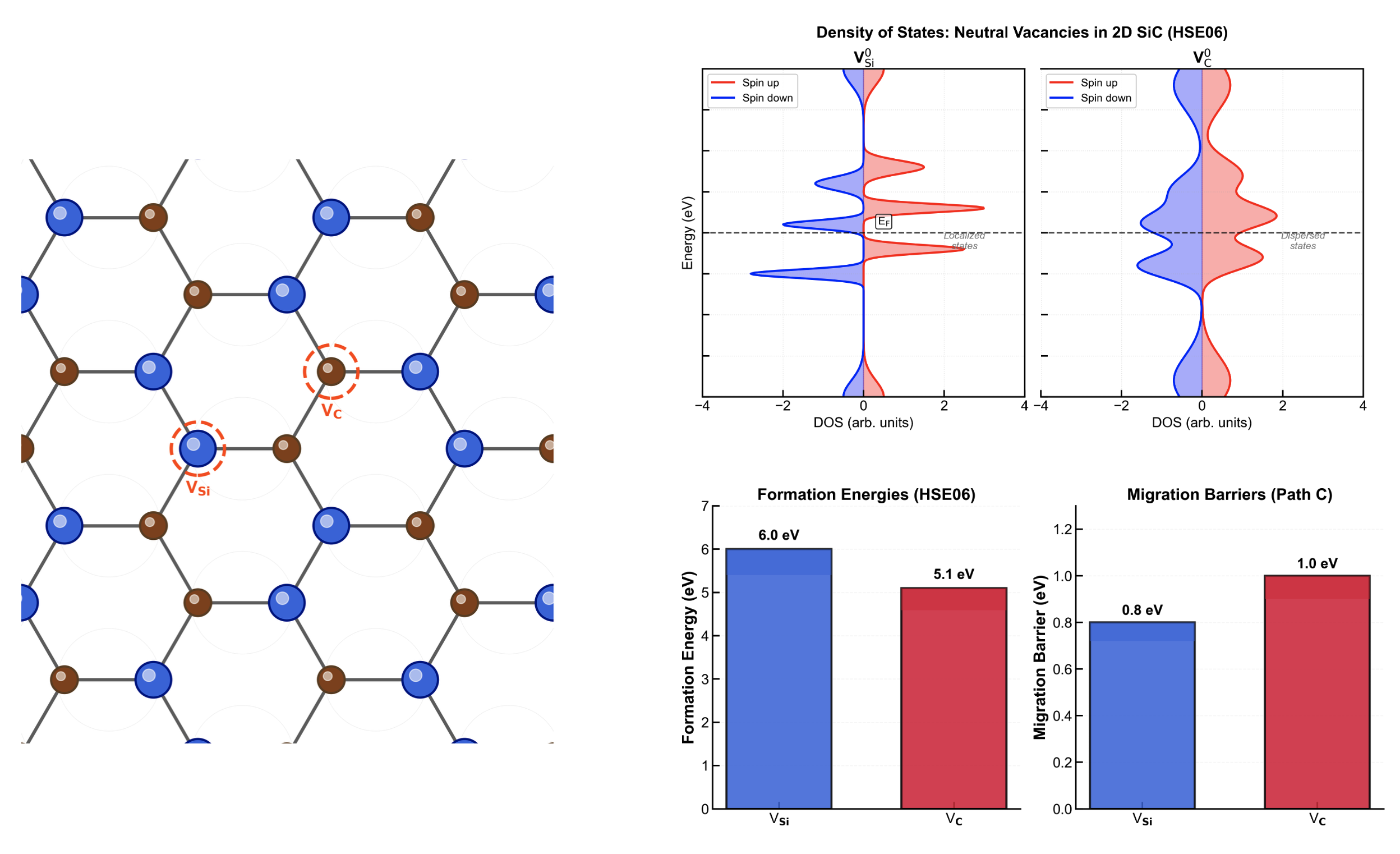}
		\caption{Graphical Abstract}
		\label{fig:enter-label}
	\end{figure}
\end{document}